\documentclass[prc,multicol,epsf,aps]{revtex4}

\usepackage{graphicx}
\usepackage{amssymb}
\usepackage{amsmath}

\begin{document}  
\draft
 
 
 
\title { Conversion of electron spectrum associated with fission into the antineutrino spectrum }

\author {Petr Vogel}

\address {Kellogg Radiation Lab. 106-38, California Institute 
of Technology, Pasadena, CA 91125, USA \\
and 
Institute de Physique Nucl\'{e}aire, F-91406 Orsay cedex, France} 
 
\date{\today}  
 




\vspace{0.4cm} 
 
\begin{abstract} 
The accuracy of the procedure that converts the experimentally determined electron spectrum
associated with fission of the nuclear fuels $^{235}$U, $^{239}$Pu, $^{241}$Pu, and $^{238}$U
into the $\bar{\nu}_e$ spectrum is examined. By using calculated sets of mutually consistent
spectra it is shown that the conversion procedure can result in a small $\sim$1\% error provided
several conditions are met. Chief among them are the requirements that the average nuclear
charge $\langle Z \rangle$ as a function of the $\beta$ decay endpoint energy is independently
known and that the   $\bar{\nu}_e$ spectrum is binned into bins that are several times larger
than the width of the slices used to fit the electron spectrum.
\end{abstract}  

\maketitle
 
 
 
\section{Introduction} 

Good knowledge of the $\bar{\nu}_e$ spectra of nuclear power reactors is an important ingredient 
of the study of neutrino oscillations with reactors \cite{RMP}. The next generation of 
oscillation experiments
(e.g. Daya-Bay \cite{Daya} and Double Chooz \cite{DCH}) aim at an unprecendented  accuracy
in their search for the mixing angle $\theta_{13}$. Even though the multi-detector scheme
used in them will greatly reduce the dependence on knowledge of the reactor spectrum,
a better understanding of it, and the corresponding error bars, would be clearly beneficial.
Nuclear reactor monitoring with neutrinos is another 
potential application of the reactor $\bar{\nu}_e$ detection
\cite{Cribier}. Again, good understanding of the spectrum and its errors is a necessary
condition for its success. 

The power in nuclear reactors originates in the (time dependent) contribution of fission
of four nuclear fuels: $^{235}$U, $^{239}$Pu, $^{241}$Pu, and $^{238}$U. During a typical
fuel cycle the dominant contribution of $^{235}$U at the beginning of the cycle
slowly decreases, and the contribution of the reactor produced nuclei  $^{239}$Pu
and $^{241}$Pu increases. While $^{238}$U represents $\sim$ 97\% of the fresh fuel rods,
its fission, caused only by the fast neutrons,  contributes only about 10\% of the reactor
power and changes little during the refueling cycle. 
In order to determine the reactor $\bar{\nu}_e$ spectrum, and its time development, 
one has to know, therefore, the corresponding 
$\bar{\nu}_e$ spectra associated with  the delayed $\beta$ decay 
of the neutron rich fission fragments of these four fuels.

There are two principal approaches to the determination of the $\bar{\nu}_e$ spectra
of the individual nuclear fuels. One of them combines knowledge of the fission yields,
i.e., the probabilities of populating the individual fission fragments characterized by their
mass $A$ and charge $Z$, with the knowledge of their, often complex,
 $\beta$ decay. The $\bar{\nu}_e$ spectrum is then the sum of the contributions of all
 $\beta$ branches of all fission fragments. The main source of uncertainty of this method is the fact
 that the $\beta$ decay characteristics of some fission fragments
 are unknown (or poorly determined),
 particularly for those with very short lifetimes and hence high $Q$ values.
 It is then necessary to use various nuclear models in order to characterize the
 $\beta$ decay of these ``unknown" nuclei.
 
 The other approach is based on the experimentally determined {\it electron} spectrum
 associated with fission of the individual fuels which is  ``converted" into
 the $\bar{\nu}_e$ spectrum. The resulting  spectrum uncertainty is then a combination of 
 the experimental uncertainties of the electron spectrum and the uncertainty of
 the conversion procedure. The accuracy of the conversion procedure is the topic of
 the present work.
 
 The electron spectra  corresponding to the thermal neutron fission of
 $^{235}$U, $^{239}$Pu and $^{241}$Pu were measured in a series of experiments
 using the beta spectrometer BILL at the ILL High Flux Reactor in Grenoble
 \cite{Schreck} in 1982-1989.  In the quoted papers the electron spectra were
 converted into the $\bar{\nu}_e$ spectra. These spectra, supplanted by the calculated spectrum
 of $^{238}$U (typically from Ref. \cite{Vog}), were used in the analysis of essentially all
 oscillation experiments performed so far. At short distances from the reactor they
 agree quite well with the measured $\bar{\nu}_e$ signal \cite{RMP}.
 
 The conversion procedure used in Refs. \cite{Schreck} is only briefly described there.
 The corresponding error is estimated to be 3-4\% with a weak energy dependence.
 Since the statistical and normalization errors of the electron spectra usually
 exceeded the estimated conversion uncertainty, its precise value was less important so far.
 Here, I wish to return to this issue, two decades later, in order to determine the
 uncertainty in the conversion procedure more firmly, and study its possible
 optimization.
 
 \section{Conversion for one nucleus}
 
 Throughout this work it is assumed that all relevant $\beta$ decays have the allowed shape.
 The possible effects associated with the unique first forbidden decays (and higher order
 forbidden decays) are neglected. Thus, the electron spectrum is assumed to be of the
 shape
 \begin{equation}
 N(E_e) = k(E_0, Z) E_e p_e (E_0 - E_e)^2 F(Z+1,E_e) ~,
 \label{eq:allowed} 
 \end{equation}
 where $k(E_0, Z)$ is the normalization constant, 
 $E_e$ and $p_e$ are the full electron energy and momentum, and $F(Z+1,E_e)$ 
 describes the Coulomb effect on the outgoing electron.
 The $\bar{\nu}_e$ spectrum as a function of the antineutrino energy
 $E_{\nu}$ is obtained from eq.(\ref{eq:allowed}) by substituting 
 $E_{\nu} = E_0 - E_e$ in the above formula.
 
 \begin{figure}[htb]
\begin{center}
\includegraphics[width=0.65\textwidth,clip=true]{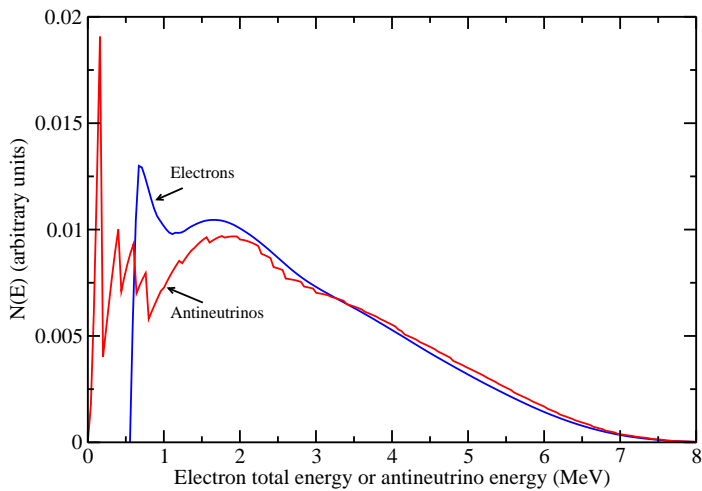}
  \caption{{\small (Color online) Example of complex electron and corresponding $\bar{\nu}_e$
  spectra. This example is for 
  a hypothetical $\beta$ decay of a $Z = 45$ nucleus
  with 40 branches. The endpoints are randomly
  distributed, with  average spacing of 200 keV, and the largest kinetic 
  energy of the electrons is 8 MeV. The branching ratios
  are also randomized.}}
 \label{fig:40l.sp}
 \end{center}
 \end{figure}

 For a single branch decay, therefore, the conversion procedure is a trivial one.
 Note the basic differences between the electron and  $\bar{\nu}_e$ spectrum
 in that case. There is a finite probability of emission of an electron with 
 vanishing momentum. Thus, the electron spectrum begins at a finite value 
 at $E_e = m_e ~(p_e = 0)$,
 while the $\bar{\nu}_e$ spectrum vanishes for $E_{\nu} = 0$. On the other hand,
 near the endpoint the electron spectrum vanishes as $E_e \rightarrow E_0$, while
 the $\bar{\nu}_e$ spectrum approaches a finite value as $E_{\nu} \rightarrow E_0 - m_e$.
 Moreover, since electrons are attracted to the positively charged nucleus the
 electron spectrum reaches its maximum at lower energies than the $\bar{\nu}_e$
 spectrum.
 
An illustrative example of a complex spectrum is shown in Figure \ref{fig:40l.sp}. Note that the 
full electron energy is used on the x-axis; hence the electron spectrum begins at $m_e$.
Due to the differences listed above the smooth electron spectrum is associated with a 
jagged $\bar{\nu}_e$ spectrum.  That is most noticeable at low energies, while at higher
energies the two spectra are quite similar. (The very upper edge is not shown. Naturally
the electron spectrum extends $m_e$ further than the $\bar{\nu}_e$ spectrum.)  

\begin{figure}[htb]
\begin{center}
\includegraphics[width=0.65\textwidth,clip=true]{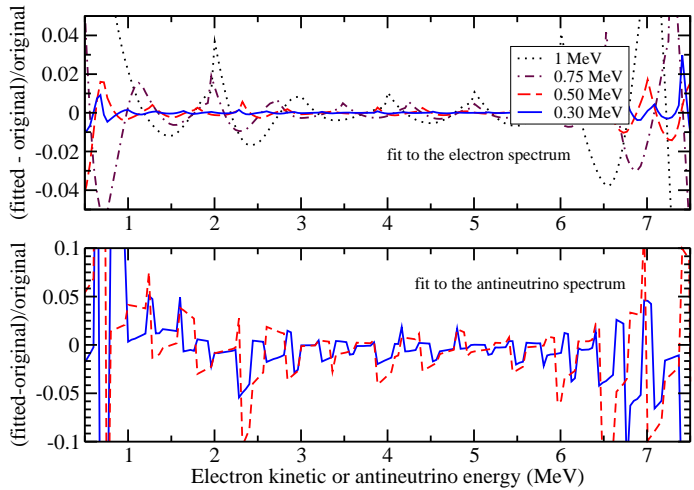}
  \caption{{\small (Color online) The fit to the complex spectra in Figure \ref{fig:40l.sp}. 
  In the upper panel the smooth
  electron spectrum is fitted. The size of the slices and the corresponding notation is indicated.
  The plotted quantity is the normalized deviation of the fitted spectrum from the original one.
  In the lower panel the endpoints and intensities of the fitted branches from the electron spectrum
  are used to reconstruct the $\bar{\nu}_e$ spectrum. Its normalized deviation from the
  true original $\bar{\nu}_e$ spectrum is plotted for two slice intervals
  (300 keV - full line and 500 keV - dashed line). Note the change of scale between the panels.}}
 \label{fig:40l.fit}
 \end{center}
 \end{figure}

Let assume that the electron spectrum in Fig. \ref{fig:40l.sp} has been 
experimentally determined,
but the endpoints and intensities of the individual branches remain unknown.
How can one convert the electron spectrum into the $\bar{\nu}_e$  one? The
procedure used in Refs. \cite{Schreck} is sensible and practical. The electron spectrum
is divided into a number $n$ of slices. In  \cite{Schreck} these slices are all of equal width,
and that choice again is simplest and economical, since it minimizes the number of
free parameters. Then, starting with the highest energy slice, one assumes the
allowed shape, Eq.(\ref{eq:allowed}), and fits for the corresponding endpoint and intensity.
Once these parameters were determined, the resulting branch spectrum is extended down
to $E_e = m_e$ and subtracted from the full spectrum. The procedure is  then repeated for the
next slice, etc., until all $n$ slices (or assumed branches) are characterized by their
endpoints and intensities. The free parameter $n$, the number of assumed branches, is then
optimized so that the true original 
experimental electron spectrum, and the fitted one, are as close to each
other as possible.

Since the electron spectrum is always a smooth function of energy, the procedure works well,
as demonstrated in the upper panel of Fig. \ref{fig:40l.fit}.  For slices of 0.3 or 0.5 MeV width
the maximum deviation is less than 1\%, except for the lowest and highest electron energies. 
(Since the spectrum is assumed to be measured
in a finite number of points, every 50 keV in this example, the number of slices $n$ 
is constrained also from above.)
Having approximated the complex $\beta$ decay scheme by $n$ 
fictitious branches characterized by their endpoints and branching ratios, one can construct 
the corresponding $\bar{\nu}_e$ spectrum. Comparing it to the true original spectrum, one
can see that the fit is less perfect (see the lower panel in Fig.  \ref{fig:40l.fit}). 
That is not surprising.
The steps associated with true endpoints cannot be faithfully reproduced by the fit, hence
the jagged form in Fig.  \ref{fig:40l.fit}. For practical application this is less important; if the
$\bar{\nu}_e$ spectrum is binned in bins wider  than the slices of the fit, the up and down features
in Fig.  \ref{fig:40l.fit} would be smoothed out.

\section{Spectrum of fission fragments}

The fission fragments are broadly distributed in nuclear masses and charges.
For example, the calculations used in Ref. \cite{Vog} were based on 1153 different
fission fragments ranging from $Z,A$ = 23,66 to 72,172. Most of them are radioactive,
and many have complex $\beta$ decay schemes. The resulting electron and $\bar{\nu}_e$
spectra are therefore superpositions of many thousands of $\beta$ decay transitions.

\begin{figure}[htb]
\begin{center}
\includegraphics[width=0.65\textwidth,clip=true]{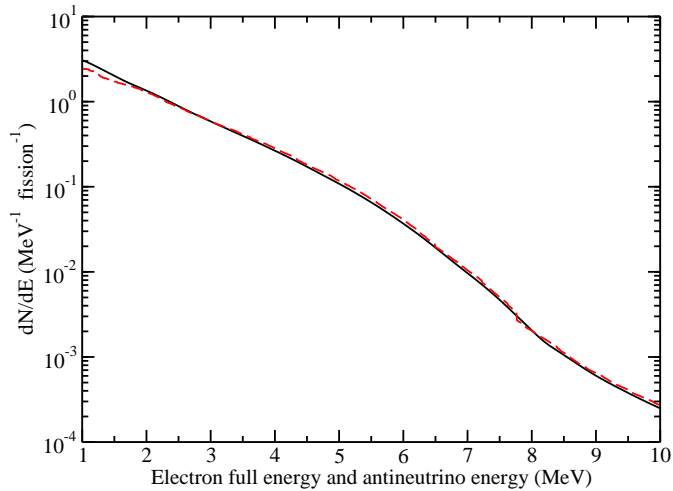}
\caption{{\small (Color online) Calculated spectra of electrons and $\bar{\nu}_e$ corresponding 
to the thermal neutron fission
of $^{235}$U at equilibrium. }}
\label{fig:u35_sp}
 \end{center}
 \end{figure}

 An example of the calculated electron and $\bar{\nu}_e$ spectra is shown in 
 Fig. \ref{fig:u35_sp}. They correspond to the thermal neutron fission of $^{235}$U
 using the data files in Ref. \cite{Vog}. Clearly, within the
 rather coarse scale of the figure, the two spectra are essentially identical; thus
 the conversion procedure should work well. Both spectra, calculated with a fine step
 of 10 keV, appear to be smooth. In the case of $\bar{\nu}_e$ this is so because 
 so many $\beta$ decays contribute that the
 step at the endpoint of each of them is not noticeable.
 
 The equilibrium spectrum in Fig. \ref{fig:u35_sp} is calculated using
 \begin{equation}
\frac{dN}{dE_e} = \Sigma_{A,Z} Y(Z,A) \Sigma_i b_i E_e p_e (E_0^i - E_e)^2 F(Z+1,E_e) ~,
\label{eq:fiss_sp}
\end{equation}
where $Y(Z,A)$ are the cummulative fission yields, and  $b_i$ are the $\beta$ decay branching ratios
of the $i$th branch
of the fragment $Z,A$ with endpoints $E_0^i$. The $\bar{\nu}_e$ spectrum is obtained from
the same formula by the substitution $E_{\nu} = E_0^i - E_e$ for every branch.

As stated above, the allowed shape is assumed for all $\beta$ branches. This assumption
should not affect the conclusions
of this work, which is based on the consistent comparison 
of the converted $\bar{\nu}_e$ spectrum with the
calculated spectrum using Eq. (\ref{eq:fiss_sp}). On the other hand, the actual spectrum
associated with fission will clearly contain a number of forbidden $\beta$ transitions.
Among them the nonunique first forbidden ones have usually allowed shapes anyway, and
the higher order forbidden decays could be safely neglected. It is difficult to estimate
the error associated with the neglect of the known shape factor of the unique first 
forbidded $\beta$ decays, but it is not expected to be significant, since the converted
 $\bar{\nu}_e$ spectra agree well with measurements at close distances from the reactor
\cite{RMP}.

Now, let us use the conversion procedure described above and convert the electron spectrum,
assumed to be known accurately, into the  $\bar{\nu}_e$ spectrum. When the smooth electron
spectrum is divided into $n$ slices of sufficiently narrow width, each slice can be well 
reproduced as a piece of the allowed spectrum, Eq.(\ref{eq:allowed}). There is only one 
parameter to fit for each slice, the endpoint $E_0$, while the normalization 
is determined automatically. 
The dependence on the nuclear charge $Z$ is rather weak, a good fit is obtained
with a constant $Z$.

\begin{figure}[htb]
\begin{center}
\includegraphics[width=0.65\textwidth,clip=true]{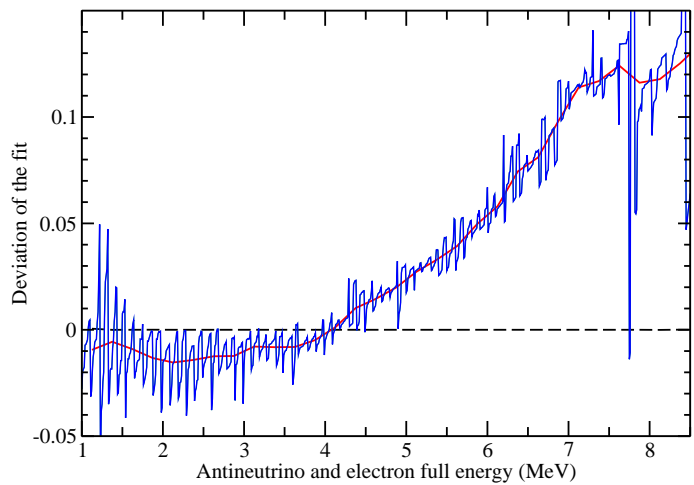}
\caption{ {\small (Color online) 
Deviations of the fit for the electron (dashed line; deviations very small and not 
visible in the plot) and $\bar{\nu}_e$ spectra,
calculated with a constant $Z$ = 47.
The fit with 100 keV slices is the jagged line; interpolating the antineutrino  spectrum
over  250 keV wide bins the smoother 
line results. The spectra are for the  thermal neutron fission of $^{235}$U at equilibrium
as in Fig. \ref{fig:u35_sp}.}}
\label{fig:z47}
 \end{center}
 \end{figure} 
 
 \begin{figure}[htb]
\begin{center}
\includegraphics[width=0.65\textwidth,clip=true]{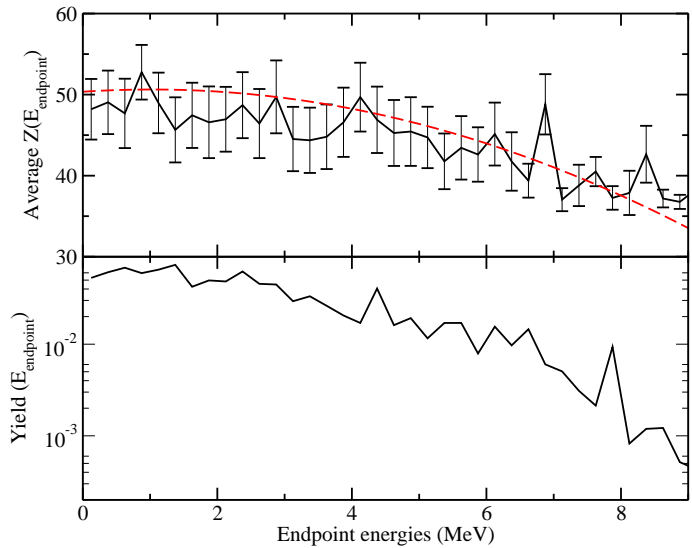}
\caption{ {\small (Color online) 
Average nuclear charge $\langle Z \rangle$ as a function of the $\beta$ decay
endpoint energy (in 0.25 MeV steps) for the $^{235}$U fission by thermal neutrons.
(For the definition of $\langle Z \rangle$ and its variance see text). 
The quadratic polynomial $Z(E_0)$ that gave the best fit to the $\bar{\nu}_e$ spectrum
is shown by the dashed line. 
The lower panel shows,
in log scale, the fission yield as a function of $E_{endpoint}$.}}
\label{fig:zyav}
 \end{center}
 \end{figure}

Having the set of $n$ endpoints and branching ratios, one can 
again reconstruct the   $\bar{\nu}_e$ spectrum and compare it
with the $\bar{\nu}_e$ spectrum derived from eq. (\ref{eq:allowed}).
There one encounters two problems. First, instead of many thousands of  branches,
there is now a much smaller number
 $n$ of branches. Hence, the steps at the endpoint of
each branch become more pronounced in comparison with the true $\bar{\nu}_e$
spectrum. Second, the choice of the nuclear charge $Z$ becomes important, unlike in the
fit of the electron spectrum. These features are illustrated in Fig.\ref{fig:z47}. There,
the fit was performed with a constant Z = 92/2+1 = 47, with slices 100 keV wide which were
also rebinned to 250 keV wide bins. The electron spectrum, with or without binning,
is reproduced perfectly, with deviations of $\sim 10^{-4}$. On the other hand, the
 $\bar{\nu}_e$ spectrum is not fitted very well. Without binning one can clearly see
 the steps caused by the fit. They disappear with binning, but the overall fit, particularly
 at higher energies, is not very good. That is caused by the choice of the fixed
 nuclear charge $Z$.

It turns out that different $Z$ values contribute to different 
parts of the spectrum. This is shown in Fig. \ref{fig:zyav}. There, for each bin 
with the endpoint $E_0$
the quantity $\langle Z \rangle$ is calculated as
\begin{equation}
\langle Z \rangle (E_0) = \frac{\Sigma_{A,Z}  Y(Z,A) \Sigma_i b_i (E_0^i) Z }
{\Sigma_{A,Z}  Y(Z,A) \Sigma_i b_i (E_0^i)} ~~,~~
\Delta<Z> = (\langle Z^2 \rangle - \langle Z \rangle^2)^{1/2} ~.
\label{eq:zav}
\end{equation} 
The error bars in Fig. \ref{fig:zyav} are the dispersion $\Delta<Z>$
of the $Z$ values within each bin.
One can see clearly the slope of the $\langle Z \rangle (E_0)$ function; the $\langle Z \rangle$ 
values are decreasing with increasing $E_0$. At the same time, in the lower panel of 
Fig. \ref{fig:zyav}, one can see that the total fission yield decreases very rapidly with increasing
$E_0$.

\begin{figure}[htb]
\begin{center}
\includegraphics[width=0.6\textwidth,clip=true]{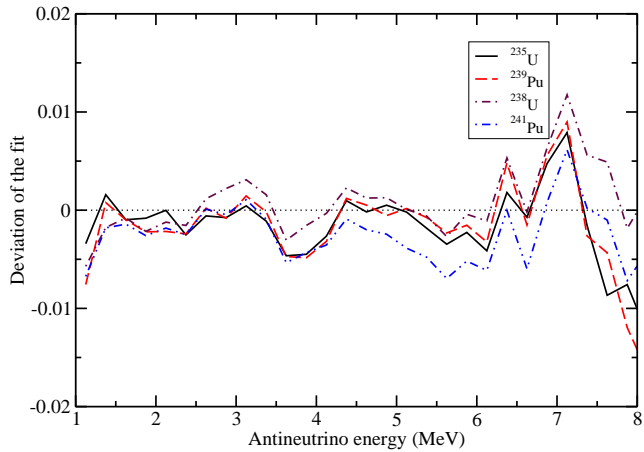}
\caption{{\small (Color online) 
Deviations of the $\bar{\nu}_e$ fits using the quadratic polynomial expression
for nuclear charge dependence on the endpoint, $Z(E_0)$ (see Table I), for the indicated
nuclear fuels. Input spectra with  10 keV intervals  were fitted in 100 keV
wide slices which were then interpolated into 250 keV
wide bins. }}
\label{fig:fitdev}
 \end{center}
 \end{figure}

The smooth dashed line in the upper panel of Fig. \ref{fig:zyav} shows the average nuclear charge
$Z$ giving fit to the $\bar{\nu}_e$ spectrum that agrees with the original spectrum within 1\% in the
relevant interval of energies. It is represented by a quadratic polynomial in $E_0$ as in some
of the Refs. \cite{Schreck}. The dashed line is above the corresponding $<Z>$ due to the fact
that in the eq.(\ref{eq:allowed}) $Z+1$ should be used. The quadratic polynomial fits of the
form $Z = a + b \times E_0 + c \times E_0^2$ were used for all four nuclear fuels, resulting in
deviation that does not exceed $\sim$1\% as shown in Fig. \ref{fig:fitdev}. The corresponding
parameters $a,b,c$ are listed in Table I. The procedures (width of the slices and
interpolating bins) used in the fits shown in Fig. \ref{fig:fitdev} appear to be close to the optimum
required for  the precision conversion procedure.

The calculated electron and $\bar{\nu}_e$ spectra used for testing the accuracy of the conversion
procedure were mutually consistent, but did not contain the small but nonnegligible QED
and weak magnetism corrections.
In conversion using the experimentally measured electron spectra these corrections should be 
included. The corresponding formulae can be found e.g. in Refs. \cite{Vogel84,Fayans85,Kurylov03}.
\begin{table}
\begin{center}
\caption{{\small Parameters of the polynomial fit 
$\langle Z \rangle = a + b\times E_0 + c \times E_0^2$ of the average
nuclear charge as a function of the endpoint energy for the indicated
isotopes }}
\medskip
\begin{tabular}{|r|rrr|}
\hline
nucleus & $a$ & $b$ & $c$ \\
\hline
$^{235}$U & 50.0 & 0.825 & -0.269 \\
$^{239}$Pu &  50.0 & 0.825 & -0.250 \\
$^{241}$Pu & 50.0 & 0.825 & -0.220 \\
$^{238}$U & 50.5 & 0.825 & -0.220 \\
\hline
\end{tabular}
\end{center}
\label{tab:par}
\end{table}

\section{Conclusion}

Using the mutually consistent sets of the realistic electron and $\bar{\nu}_e$ spectra
corresponding to the fission of the nuclear fuels $^{235}$U, $^{239}$Pu, $^{241}$Pu, 
and $^{238}$U I have shown that the electron spectrum can be converted into the
$\bar{\nu}_e$ spectrum with an error that does not exceed $\sim$1\% in the energy
interval 1 - 8 MeV. However, such accurate conversion can be obtained only if several
conditions are met:
\begin{itemize}
\item The slices into which the electron spectrum is divided are sufficiently fine.
\item  The converted $\bar{\nu}_e$ spectrum is smoothed out by binning in bins
that are several times larger than the width of the original slices.
\item The optimum average nuclear charge $\langle Z \rangle$ is independently
known as a function of the endpoint energy $E_0$.
\end{itemize}

 Not all these criteria were fulfilled in Refs. \cite{Schreck}, hence the larger uncertainties 
 associated with the conversion procedure assigned there cannot be safely reduced. 
 The electron spectra in \cite{Schreck} were published in 250 keV wide bins. The number
 of bins and the number of slices for the conversion procedure were the same (or nearly
 so). The smooth electron spectra were expressed in the Fermi-Kurie representation 
 which made the corresponding fits possible. And the $\bar{\nu}_e$ spectra were not rebinned,
 thus the steps at the edges of slices were not smoothed out. However, as already 
 pointed out above, the statistical and systematic errors of the 
 measured electron spectra mostly exceeded
 the errors of the conversion procedure, so even if it would be somehow possible to
 minimize the uncertainty of the conversion procedure, it would affect the spectra
 of Refs. \cite{Schreck} only minimally. They remain the best existing source of the fission
 associated  $\bar{\nu}_e$ spectra.

 The main purpose of the present work is an examination of the conversion procedure per se.
 While there are no immediate plans to repeat and/or improve the results of the 
 experiments in Refs. \cite{Schreck}, there exist measurements of the $\beta$ spectra
 of many individual neutron rich fission fragments (see, e.g. \cite{Allek,Tengb}). These, in turn,
 can be converted into the $\bar{\nu}_e$ spectra using the procedure described here,
 and combined with the fission yields to obtain the full $\bar{\nu}_e$ spectrum.
 Similar measurements of individual short lived (and high $Q$ value) fission
 fragments can be performed with the modern isotope separation facilities. Thus,
 significant improvements of the determination of the reactor $\bar{\nu}_e$ spectra
 is possible. The present work is part of that effort.

\section{Acknowledgment}

The work reported here was supported in part by the US Department of Energy contract
DE-FG02-05ER41361. The author deeply appreciates the hospitality of Prof. Cristina Volpe
and the Institut de Physique Nucl\'{e}aire, Orsay, where part of this research was performed.

 
 {}


\begin{thebibliography}{}
 
 \bibitem{RMP} C. Bemporad, G. Gratta and P. Vogel, Rev.  Mod. Phys. {\bf 74},
 297 (2002).
 
 \bibitem{Daya} Daya-Bay collaboration, arXiv: hep-ex/0701029.
 
 \bibitem{DCH} F. Ardellier {\it et al.}, arXiv:hep-ex/0606025.
 
 \bibitem{Cribier} M. Cribier, arXiv:0704.0548; arXiv:0704:0891.
 
 \bibitem{Schreck} F. von Feilitzsch, A.A.Hahn and K. Schreckenbach, Phys. Lett. {\bf 118B},
 162 (1982); K. Schreckenbach, G. Colvin, W. Gelletly and F. von Feilitzsch,
 Phys. Lett. {\bf 160B}, 325 (1985); A. A. Hahn {\it et al.} Phys. Lett. {\bf 218B}, 365 (1989).
 
 \bibitem{Vog} B. R. Davis, P. Vogel, F. M. Mann and R. E. Schenter,
 Phys. Rev C{\bf 19}, 2259 (1979); P. Vogel,  G. K. Schenter, F. M. Mann and
 R. E. Schenter, Phys. Rev. C{\bf 24}, 1543 (1981).
 
\bibitem{Vogel84} P. Vogel, Phys. Rev. D{\bf 29}, 1918 (1984).

\bibitem{Fayans85} S. A. Fayans, Sov. J. Nucl. Phys. {\bf 42}, 590 (1985).

\bibitem{Kurylov03} A. Kurylov, M. J. Ramsey-Musolf and P. Vogel, Phys. Rev. C{\bf 67}, 035502 (2003). 

\bibitem{Allek} K. Aleklett, G. Nyman, and G. Rudstam, Nucl. Phys.
{\bf A246}, 425 (1977).

\bibitem{Tengb} O. Tegblad {\it et al.}, Nucl. Phys. {\bf A503}, 136 (1989).

\end{thebibliography}
\end{document}